\documentclass[a4paper,11pt]{article}
\usepackage{pos}

\title{Critical exponents for a spin-charge flip symmetric fixed point in 2+1d with massless Dirac fermions}
\ShortTitle{Spin-charge Flip Symmetric Fixed Point}

\author*[1]{Emilie Huffman}\note{Work done in collaboration with Shailesh Chandrasekharan, Ribhu Kaul, and Hanqing Liu}

\affiliation{Perimeter Institute for Theoretical Physics,\\
  31 Caroline St N, Waterloo, Ontario, Canada}

\emailAdd{ehuffman@perimeterinstitute.ca}

\abstract{In the Hamiltonian picture, free spin-$1/2$ Dirac fermions on a bipartite lattice have an $O(4)$ (spin-charge) symmetry. Here we construct an interacting lattice model with an interaction $V$, which is similar to the Hubbard interaction but preserves the spin-charge flip symmetry. By tuning the coupling $V$, we show that we can study the phase transition between the massless fermion phase at small-$V$ and a massive fermion phase at large-$V$. We construct a fermion bag algorithm to study this phase transition and find evidence for it to be second order. Numerical study shows that the universality class of the transition is different from the one studied earlier involving the Hubbard coupling $U$. Here we obtain some critical exponents using lattices up to $L=48$.
}

\FullConference{%
 The 38th International Symposium on Lattice Field Theory, LATTICE2021
  26th-30th July, 2021
  Zoom/Gather@Massachusetts Institute of Technology
}

\begin{document}
\maketitle

\section{Introduction}
Fixed points that involve relativistic fermions are still a challenge to study nonperturbatively, especially in two spatial dimensions and higher, with sign problems preventing their numerical study via quantum Monte Carlo. Additionally, even when sign-problem-free methods to study fermionic fixed points are available, there is a second challenge involving the expense and efficiency of such calculations. While there are techniques available for nonperturbatively studying spin systems of tens of thousands of sites \cite{sandvik}, fermionic studies are more often limited to hundreds of sites instead, limiting what can be learned from finite-size scaling. This is a particularly important obstacle to clear given that fermions are such a fundamental part of nature, and their interactions are thus of interest across the subfields of physics, from particle physics to quantum information to condensed matter physics. It would thus be useful to find new and easy-to-study ways of representing simple fermionic fixed points, in the vein of de-sign-er Hamiltonians. \cite{Kaul:2012zv}

With this in mind, we introduce a simple class of $N$-flavor Dirac fermion models that possess the same flavor symmetry as free fermions, and yet also contain interactions. Moreover, these models are sign-problem-free and straightforwardly simulatable with the Hamiltonian fermion bag method, with the $N=1$ case in $2+1d$ already studied extensively \cite{Huffman:2017swn,Huffman:2019efk}. After discussing the class of models generally, we focus specifically on the $N=2$ case in $2+1d$, which resembles the microscopic model for the chiral Heisenberg universality class (Hubbard model on a $\pi$-flux lattice), but contains an additional spin-charge-flip symmetry, making its flavor symmetry $O(4)$, which is a symmetry of physical interest \cite{assaad}. We present numerical results using the Hamiltonan fermion bag method to explore the phase transition as a function of its coupling $V$. This proceedings provides a more detailed discussion of the numerics in \cite{Liu:2021otq}, along with some additional numerical results to support our conclusions there, while more details on a corresponding continuum analysis for the fixed point may be found in \cite{Liu:2021opm}.

\section{Model and Symmetries}
We begin with an $N$-flavor extension of the $t$-$V$ model in $2+1d$, given by
\begin{equation}
    H = -\prod_{a=1}^N \left[t\sum_{x,d} \eta_{x,\hat{d}}\left(c^\dagger_{x,a} c_{x+\hat{d},a} + c^\dagger_{x+\hat{d},a} c_{x,a}\right) - V\sum_{x,d} \left(n_{x,a}-\frac{1}{2}\right)\left(n_{x+\hat{d},a} - \frac{1}{2}\right) + \frac{t^2}{V}\right],
    \label{initfamily}
\end{equation}
where $x=(x_1,x_2)$ is a two-dimensional vector on a square spatial lattice, and $c^\dagger_x$ and $c_x$ are creation and annihilation operators respectively for an electron on site $x$ with flavor $a$. The $\eta_{x,\hat{d}}$ factors induce a $\pi$-flux on the square lattice by their definitions $\eta_{x,\hat{d}_x}=1$ and $\eta_{x,\hat{d}_y} = (-1)^{x_1}$, where $\hat{d}_x$ and $\hat{d}_y$ are the unit vectors in the $x$- and $y$-directions, and $x_1$ is the $x$-component of site $x$. While we have narrowed the focus to $2+1d$, it is possible to study this model in any dimension using sign-problem-free quantum Monte Carlo \cite{Huffman:2013mla,Yao:2018}, and when $N=1$ this model is the ordinary $t$-$V$ model (with the addition of the physically unimportant constant $t^2/V$), which has been extensively studied using quantum Monte Carlo in \cite{Huffman:2017swn,Huffman:2019efk,Li:2014aoa,Wang:2014cbw}.

The addition of the constant $t^2/V$ to the factors in this family of $N$-flavor models has the effect of allowing us to also write these models in the following simple way:
\thinmuskip=0mu
\medmuskip=1mu 
\thickmuskip=2mu 
\begin{equation}
\begin{aligned}
  H &= - 
  \sum_{x,\hat{d}} \exp \Big(\kappa \eta_{x,\hat{d}} \sum_{a=1}^N (c^\dagger_{x,a} c_{x+\hat{d},a} + c^\dagger_{x+\hat{d},a} c_{x,a})\Big),
\end{aligned}
\label{expform}
\end{equation}
\thinmuskip=3mu
\medmuskip=4mu 
\thickmuskip=5mu
where we have $2\tanh \kappa / 2 = V/t$, as shown in \cite{Wang:2016,Liu:2021opm}. This form has the advantage of making some of the symmetries of these models more obvious, as it is clear that when (\ref{expform}) is expanded out, every term is a power of the free fermion Hamiltonian $H_0 = -t\sum_{x,\hat{d}} \sum_{a=1}^N \eta_{x,\hat{d}} \left(c^\dagger_{x,a} c_{x+\hat{d},a} + c^\dagger_{x+\hat{d},a} c_{x,a}\right)$, which at low energies describes $N$ four-component massless Dirac fermions in the continuum limit. In the Majorana basis it is clear that this family of models possess an $O(2N)$ flavor symmetry \cite{Huffman:2017swn}.

While the simulation methods used in these proceedings are applicable to any $N$ in this family of models, here we will focus on the $N=2$ model in particular, which we call $H_{SC}$, and is given by
\thinmuskip=0mu
\medmuskip=1mu 
\thickmuskip=2mu 
\begin{equation}
\begin{aligned}
  H_\mathrm{SC} &=-\prod_{s=\uparrow,\downarrow} \left[t\sum_{x,d} \eta_{x,\hat{d}}\left(c^\dagger_{x,s} c_{x+\hat{d},s} + c^\dagger_{x+\hat{d},s} c_{x,s}\right) - V\sum_{x,d} \left(n_{x,s}-\frac{1}{2}\right)\left(n_{x+\hat{d},s} - \frac{1}{2}\right) + \frac{t^2}{V}\right]\\
  &=- 
  \sum_{x,\hat{d}} \exp \Big(\kappa \eta_{x,\hat{d}} \sum_{s=\uparrow, \downarrow} (c^\dagger_{x,s} c_{x+\hat{d},s} + c^\dagger_{x+\hat{d},s} c_{x,s})\Big).
\label{spin-charge-model}
\end{aligned}
\end{equation}
\thinmuskip=3mu
\medmuskip=4mu 
\thickmuskip=5mu

For small values of $\kappa$, the fermion bilinear terms reproduce free fermions with $t_{ij} = \kappa \eta_{ij}$, and the higher order fermion interactions are small. Because we are looking at two flavors of fermions, (\ref{spin-charge-model}) can be written in terms of powers of free fermion terms that correspond to 2 four-component massless Dirac fermions in the continuum, and the Hamiltonian possesses $O(4)$ flavor symmetry in addition to the usual lattice symmetries and time reversal. We can break down this $O(4)$ symmetry into $SU(2)_s\times SU(2)_c\times \mathbb{Z}^{sc}_2$ symmetries \cite{Liu:2021otq}. $SU(2)_s$ is the well-known spin rotational symmetry, which is generated by $\vec{\mathcal S}_x = \frac{1}{2} c^\dagger_{xa}\vec \sigma_{ab}c_{xb}$, and $SU(2)_c$ is the well-known ``hidden'' charge symmetry, which is generated by $\vec{\mathcal{C}}_x = \frac{1}{2} (\zeta_x (c^\dagger_{x\uparrow}c^\dagger_{x\downarrow} + c_{x\downarrow}c_{x\uparrow}), -i\zeta_x (c^\dagger_{x\uparrow}c^\dagger_{x\downarrow} - c_{x \downarrow}c_{x \uparrow}), c^\dagger_{x\uparrow}c_{x\uparrow} + c^\dagger_{x\downarrow}c_{x\downarrow} - 1)$,  with $\zeta_x = \pm 1$ depending on whether the site is even or odd. Finally, $\mathbb{Z}^{sc}_2$ is a spin-charge flip symmetry defined as $\mathcal{F} c_{x\downarrow} \mathcal{F}^\dagger = \zeta_x c^\dagger_{x\downarrow}$ under which the generators of spin and charge rotations are interchanged, $\vec {\mathcal S}_x \leftrightarrow \vec {\mathcal C}_x$. Because of this symmetry between spin and charge we have given this Hamiltonian the name $H_{SC}$ to keep the spin-charge-flip symmetry $\mathbb{Z}^{sc}_2$ in mind.


We note that while the symmetries listed above are all seen in free fermions, our model $H_{SC}$ is not free and contains 4-, 6- and 8-fermion interactions. However, by maintaining the $O(4)$ symmetry we have an enhanced symmetry compared to the most often studied interaction that is added to $N=2$ free fermions: the Hubbard-$U$ term, $H_U = U \sum_x (n_{x\uparrow} - \frac{1}{2}) (n_{x\downarrow} - \frac{1}{2})$. This term still preserves the $SU(2)_s$ and $SU(2)_c$ symmetries mentioned above (which form an $SO(4)$ symmetry), but is odd under the $\mathbb{Z}^{sc}_2$ spin-charge flip operation. It is well-known that repulsive-$U$ interactions favor an antiferromagnetic (AFM) ``spin'' order parameter, and attractive-$U$ interactions favor a combined charge-density wave(CDW)/superconducting ``charge'' order parameter. Since the free fermion portion realizes a $2+1d$  Dirac dispersion, long range order sets in at a finite-$|U|$ phase transition which is described by the ``chiral Heisenberg'' Gross-Neveu-Yukawa fixed point that has been the subject of intense numerical study~\cite{Otsuka:2015iba}. Here we look into the nature of the quantum critical phenomena when we consider (\ref{spin-charge-model}) which preserves the full symmetry of the hopping problem, including the crucial $\mathbb{Z}^{sc}_2$ symmetry, which is absent in the usual Hubbard formulation. We present numerical evidence that the phase transition between Dirac semi-metal and spin-charge flip broken phase is continuous and in a new universality class.

\section{Algorithm}

\begin{figure}
    \centering
    \includegraphics{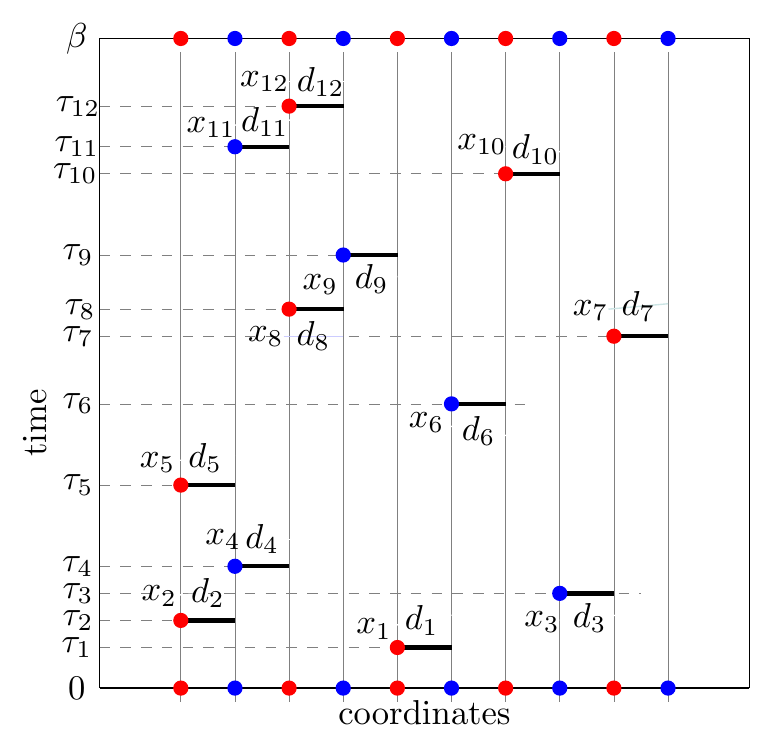}
    \caption{An example configuration from the partition function sampling space. Each bond corresponds to an $H_{x,\hat{d}}$ operator. Because $\left[H_{x,\hat{d}},H_{x',\hat{d}'}\right]=0$ when they share no sites in common, if two bonds do not touch each other in space, their corresponding operators commute. This property, which generalizes to a concept of locality, is key to the application of the Hamiltonian fermion bag method.}
    \label{fig:configuration}
\end{figure}

A nice feature of the Hamiltonians in the family given by (\ref{initfamily}) or (\ref{expform}) is that we can use an efficient fermion bag QMC algorithm to study them \cite{Huffman:2017swn,Huffman:2019efk}. We can see this immediately by comparing the Hamiltonian form from (\ref{expform}) to reference \cite{Huffman:2019efk}, and seeing that it satisfies key criteria for the algorithm to be applicable: (i) the Hamiltonian can be written as a sum of exponentiated fermionic bilinear terms, and (ii) such terms are local in terms of their degrees of freedom (in this case local in terms of the spatial lattice sites).

The following is a brief summary of how the fermion bag method works, applied to the $H_{\mathrm{SC}}$ model in particular. The partition function $Z = {\mathrm tr} e^{- H_{\mathrm{SC}}/T}$ is first expanded as
\thinmuskip=0mu
\medmuskip=1mu 
\thickmuskip=2mu 
\begin{equation}
Z = \sum_k \int \left[d\tau\right] (-1)^k {\mathrm Tr} \left[H_\mathrm{SC}(\tau_k) ... H_\mathrm{SC}(\tau_2) H_\mathrm{SC}(\tau_1)\right].
\end{equation}
\thinmuskip=3mu
\medmuskip=4mu 
\thickmuskip=5mu 
Here the notation $\int \left[d\tau\right]$ denotes time-ordered integration for times $1/T \geq \tau_k \geq \cdots \geq \tau_2 \geq \tau_1 \geq 0$. The Hamiltonian $H_\mathrm{SC}$ is not time-dependent and so $H_\mathrm{SC}(\tau) = H_\mathrm{SC}$. The expansion can be derived from the continuous-time interaction representation where $H_0 = 0$ and $H_{\rm int} = H_\mathrm{SC}$, and also resembles the stochastic series expansion \cite{Wang:2016}. The algorithm then involves exploring a configuration space made up of the terms in the expansion, which we can rewrite as
\begin{equation}
\begin{aligned}
Z &= \sum_k \sum_{\left[x,\hat{d}\right]} \int \left[d\tau\right] (-1)^k {\mathrm{Tr}}\left[H_{\mathrm{SC},x_k,\hat{d}_k}\left(\tau_k\right)...H_{\mathrm{SC},x_2,\hat{d}_2}\left(\tau_2\right) H_{\mathrm{SC},x_1,\hat{d}_1}\left(\tau_1\right)\right] \\
&H_{\mathrm{SC},x,\hat{d}}= \exp \Big(\kappa \eta_{x,\hat{d}} \sum_{s=\uparrow, \downarrow} (c^\dagger_{x,s} c_{x+\hat{d},s} + c^\dagger_{x+\hat{d},s} c_{x,s})\Big).
\end{aligned}
\end{equation}
Figure \ref{fig:configuration} gives a pictorial representation (though simplified to one spatial dimension) of a single term in the expansion: a configuration where $k=12$. Because all $H_{\mathrm{SC},x,\hat{d}}$ are expentiated fermion bilinear operators, we can then make use of the BSS formula as well as locality to compute transition probabilities as small determinants. While the method is comparatively limited compared to traditional auxiliary field methods in terms of which Hamiltonians are simulatable, it can be significantly more powerful than such methods when it is applicable \cite{Huffman:2017swn}, allowing access to large lattices with comparatively few computing hours to get critical exponents (up to $10,000$ sites in the discrete-time formulation and $4,096$ in the continuous-time formulation using on the order of a few million core-hours each).

\section{Results}

\begin{table}[b]
\begin{center}
\small
\begin{tabular}{|l||l|l|l|l|l|l|l|}
\hline
 $V/t$ & $L=12$ & $L=16$ & $L=20$ & $L=24$ & $L=28$ & $L=32$ & $L=48$ \\
 \hline
\multicolumn{8}{|c|}{$C_S$} \\
\hline
$1.480$ & 0.00226(2) & 0.00118(2) & 0.00069(2) &
0.00041(1) & 0.000279(8) & 0.000182(7) & $\qquad -$\\
\hline
$1.500$ & 0.00238(3) & 0.00126(2) & 0.00077(2) &
0.00045(1) & 0.00031(1) & 0.00023(1) &  0.000078(6)\\
\hline
$1.520$ & 0.00244(4) & 0.00134(4) & 0.00078(2) & 0.00053(2) & 0.00035(2) & 0.00027(2) & 0.00011(1)\\
\hline
$1.540$ & 0.00250(4) & 0.00140(3) & 0.00086(2) & 0.00061(3) & 0.00038(1) & 0.00033(1) & $\qquad -$ \\
\hline
$1.560$ & 0.00276(6) & 0.00148(3) & 0.00087(3) & 0.00065(2) & 0.00045(2) & 0.00037(2) & $\qquad -$ \\
\hline
$1.600$ & 0.00299(3) & 0.00172(2) & 0.00113(3) & 0.00089(4) & 0.00071(3) & 0.00068(3) & $\qquad -$ \\
\hline
\multicolumn{8}{|c|}{$C_U$} \\
\hline
$1.515$ & 0.00112(5) & 0.00060(2) & 0.00031(2) & 0.000178(7) & 0.000108(5) & $\qquad -$ & $\qquad -$ \\
\hline
$1.600$ & 0.00197(5) & 0.00110(3) & 0.00085(3) & 0.00076(5) & 0.00065(5) & 0.00063(4) & $\qquad -$ \\
\hline
\end{tabular}
\vspace{.5cm}
\caption{The QMC correlation function data that is used in the figures.}
\label{datatable2}
\end{center}
\end{table}

Using the Hamiltonian fermion bag algorithm in continuous time we compute two correlation functions of order parameters for $H_\mathrm{SC}$,
\begin{align}
    C_S  =2\langle {\cal S}^z_{x^{(0)}} {\cal S}^z_{x^{(1)}} \rangle, \quad C_U &= \langle {\cal U}^z_{x^{(0)}} {\cal U}^z_{x^{(1)}} \rangle .
    \label{qmcobs}
\end{align}
Here $C_S$ picks up the $SU(2)_s$ N\'eel order using the antiferromagnetic spin order parameter ${\cal S}^z_x = n_{x,\uparrow} - n_{x,\downarrow}$, and $C_U$ measures the breaking of the spin-charge symmetry through the local Hubbard operator ${\cal U}_x =(n_{x\uparrow}-\frac{1}{2}) (n_{x\downarrow}-\frac{1}{2})$, which is odd under $\mathbb{Z}_2^{sc}$.  The sites in the correlation functions are $x^{(0)} = (0, 0)$ and $x^{(1)} = (L/2,0)$ and we assume $L/2$ even. The QMC simulations are done at finite inverse temperature $1/T=L$, and we work with the tuning parameter $V/t=2\tanh\frac{\kappa}{2}$. Table \ref{datatable2} gives the raw data.

\begin{figure}[t]
\begin{center}
    \includegraphics[width=0.45\textwidth]{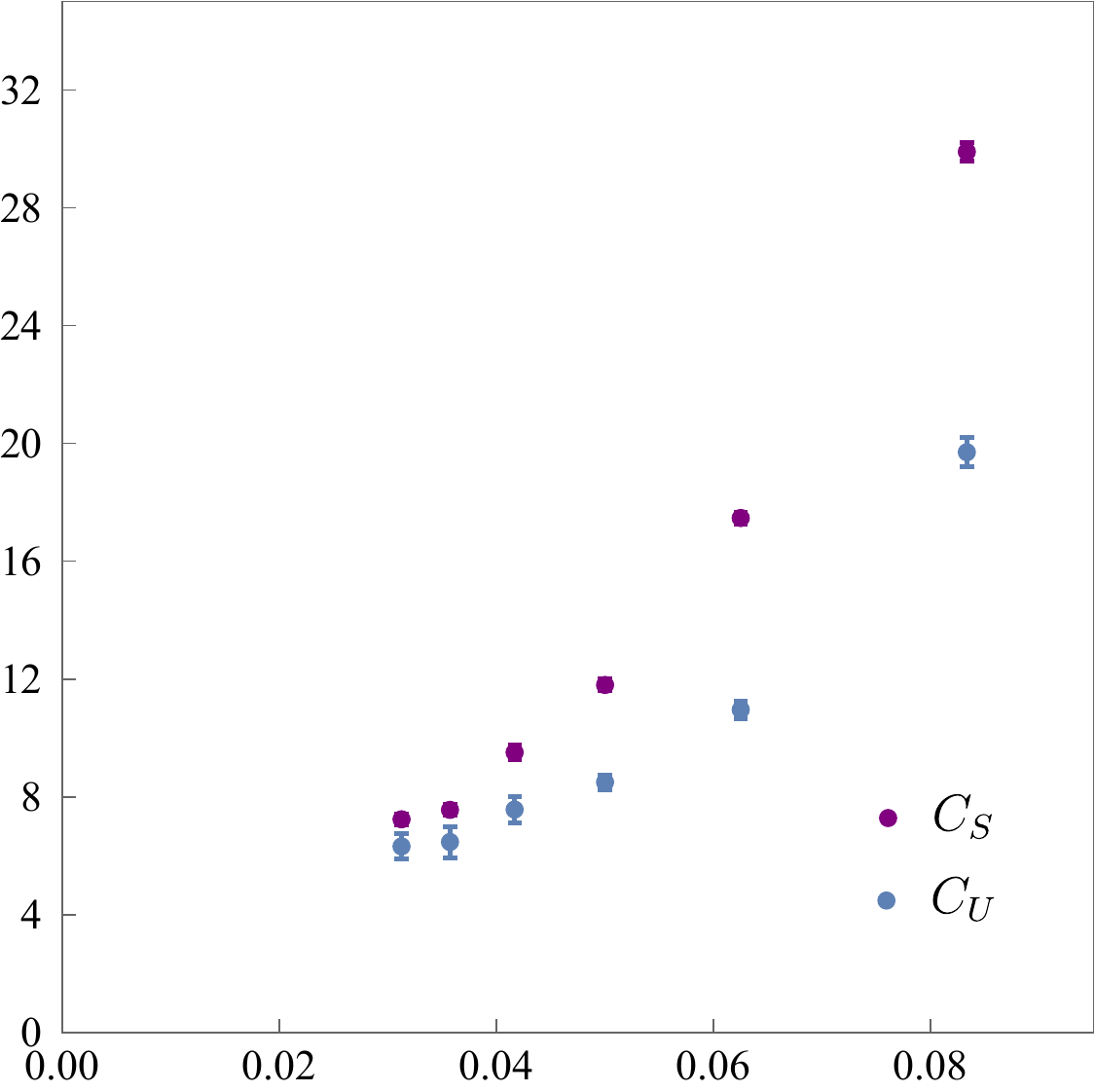}
    \end{center}
    \caption{\label{fig:massive} Characterization of the massive phase from QMC and field theory. The plot shows finite size scaling data for $C_S$ and $C_U$ using the fermion bag QMC method for a coupling constant $V/t=1.6$. Both correlation functions scale to a finite value in the thermodynamic limit indicating that the system breaks the $\mathbb{Z}_2^{sc}$ Ising symmetry as well as the SU(2) symmetry of spin and charge.
    }
\end{figure}

\begin{figure}
\begin{center}
    \includegraphics[width=0.45\textwidth]{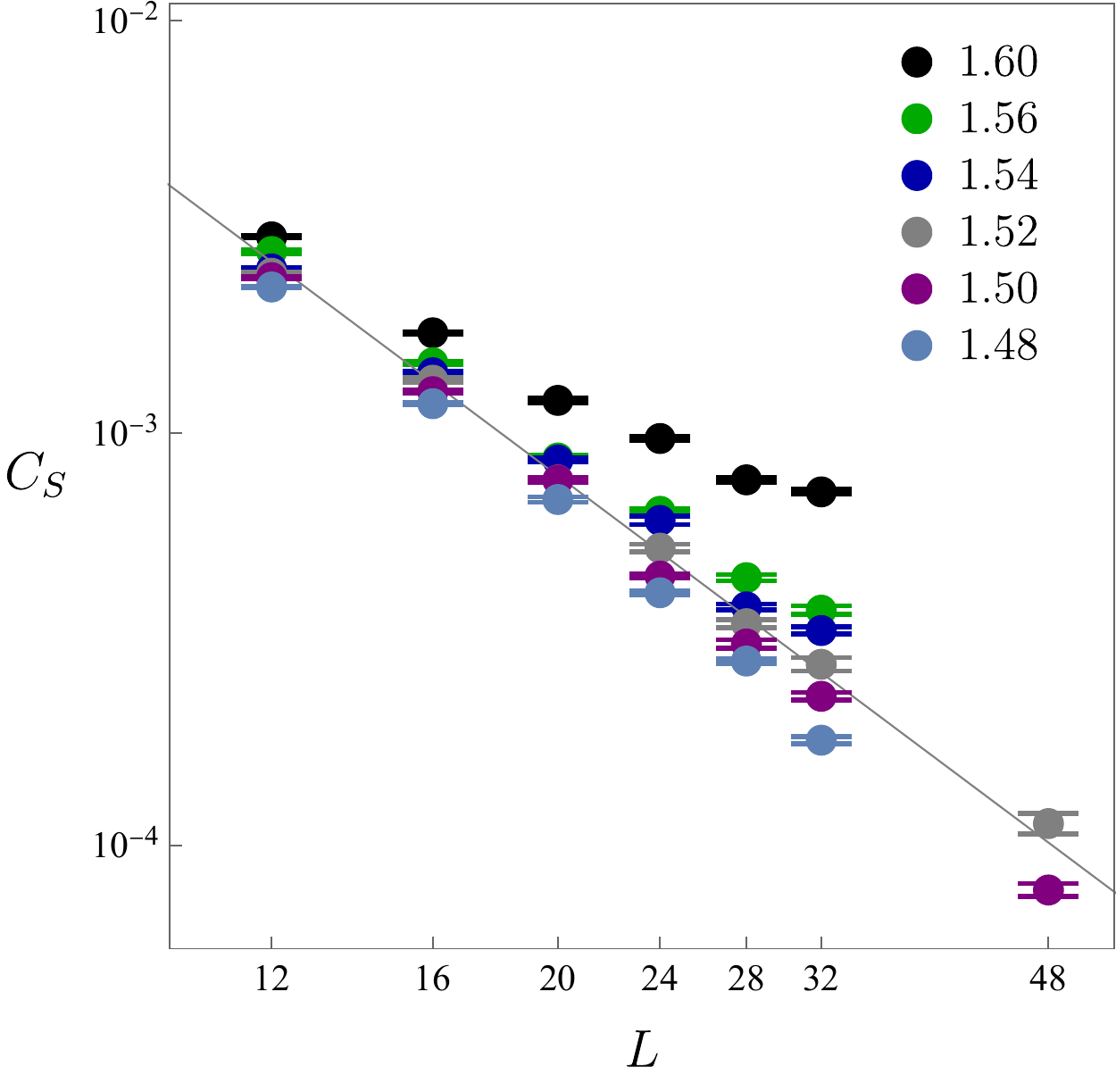}\hspace{.5cm}
     \includegraphics[width=0.475\textwidth]{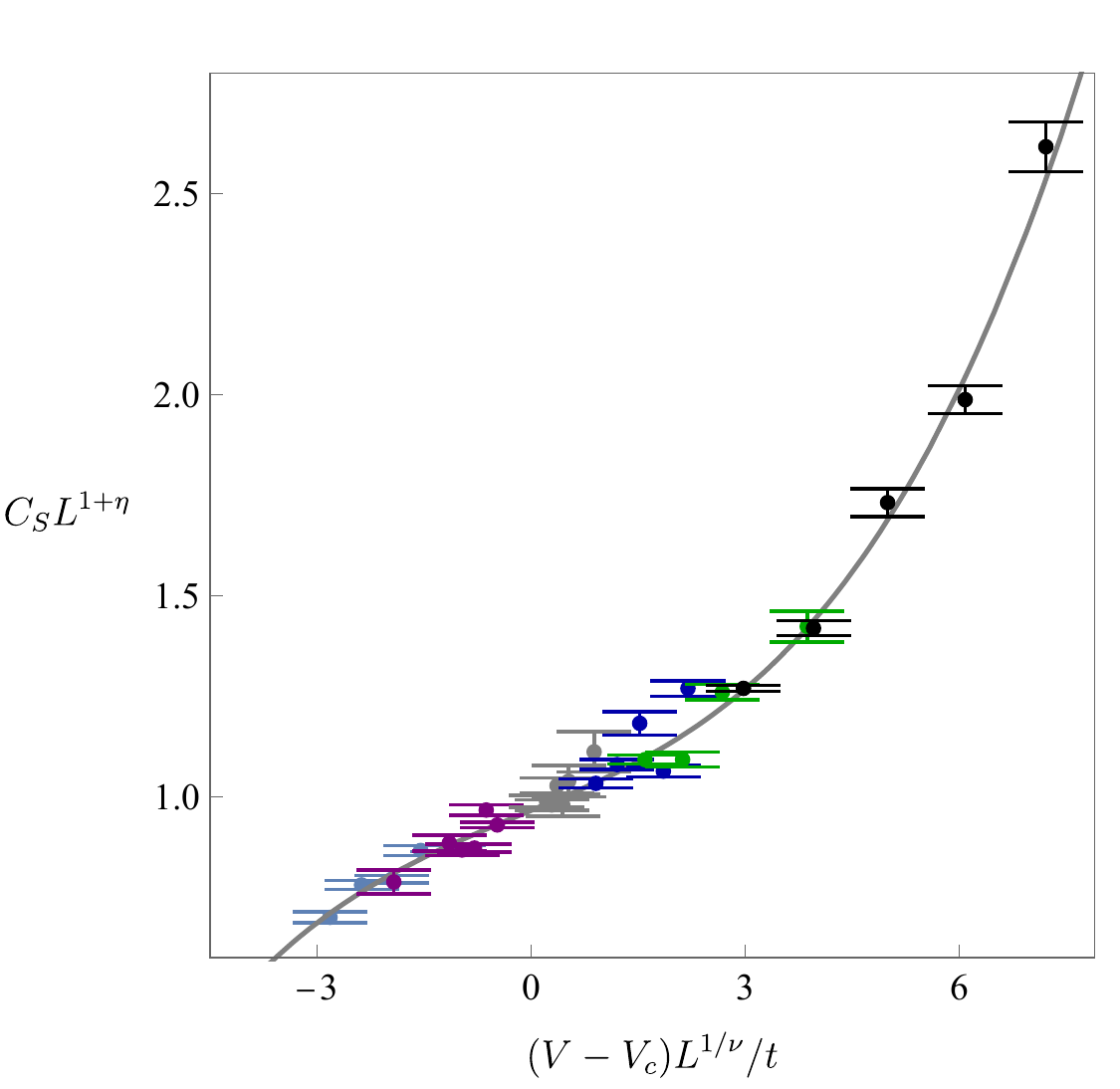}
     \end{center}
  \caption{The plot on the left shows $C_S$ as a function of $L$ on a log-log scale up to $L=48$ for various values of $V/t$. For large values of $L$ we find that $C_S$ decays to zero when $V/t = 1.48$, while it saturates to a constant when $V/t=1.60$, with a phase transition around $V_c/t \approx 1.52$ where the data fits to $C_S \approx 0.67/L^{2.25}$ (straight line in the plot). The plot on the right shows that all of the data (after dropping $L=12$) collapse to the universal scaling function discussed in the text with $\eta=1.38(6)$, $\nu=0.78(7)$, and $V_c/t=1.514(8)$, providing compelling evidence for a quantum critical point.
  }
   \label{fig:O4}
\end{figure}

\begin{figure}
\begin{center}
    \includegraphics[width=0.45\textwidth]{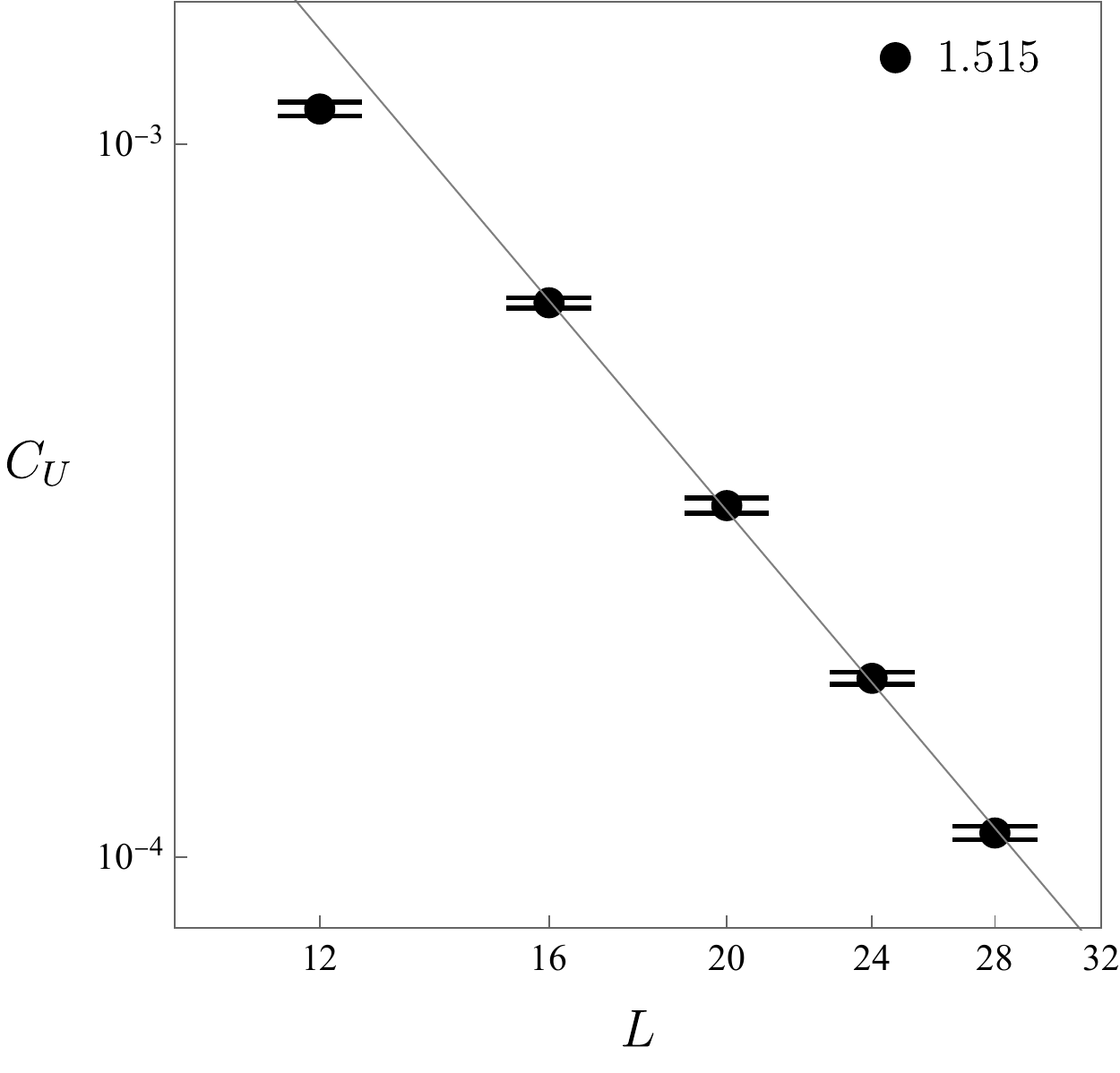}
\end{center}
  \caption{This plot shows $C_U$ at the critical coupling $V_c/t=1.515$, as found using the scaling analysis for $C_S$. All lattice sizes except for $L=12$ are used to find a power law fit $C_U = a L^p$. Here the fit gives $a=2.8(8), p=-3.05(9)$ with a $\chi^2$ of $0.2$.
  }
   \label{fig:O5}
\end{figure}

We first investigate the nature of the massive phase in our lattice model, $H_{\mathrm SC}$. We work at a coupling $V/t = 1.6$, which is deep in the massive phase. As shown in Figure \ref{fig:massive}, we find a finite value of $C_U$ in the thermodynamic limit, which indicates that the Ising symmetry, $\mathbb{Z}^{sc}_2$, is spontaneously broken. Further we find that $C_S$ also scales to a finite value in the thermodynamic limit, which implies N\'eel order and a broken $SO(4)$ symmetry (coming from the $SU(2)_s\times SU(2)_c$ symmetry). Together we interpret this to imply that the system has to spontaneously choose between the charge and the spin sector, breaking $\mathbb{Z}^{sc}_2$, and forming either a N\'eel state or a superconductor/CDW state which breaks the corresponding SU(2) symmetry.

Next, we study the nature of the phase transition between the Dirac semi-metal and the massive phase. Figure \ref{fig:O4} shows the data for $C_S$ as a function of system size $L$. For large values of $L$, there is clear evidence that $C_S$ 
converges to a nonzero constant at the coupling $V/t=1.6$ (massive phase), while it  scales to zero at the coupling $V/t=1.48$ (Dirac semimetal). A good fit to the power-law $C_s = 0.67/L^{2.25}$ for $12 \leq L \leq 48$ with a $\chi^2 = 0.95$ is found at the coupling $V/t=1.52$ as expected at a quantum critical point. A multi-parameter scaling fit of all our data except for $L=12$, to the form  $C_S = L^{-(1+\eta)} f((g-g_c)\;L^{1/\nu})$ with $f(x) = f_0 + f_1 x + f_2 x^2 + f_3 x^3$ yields $\eta=1.38(6)$, $\nu=0.78(7)$, $V_c/t=1.514(8)$, $f_0=0.96(15)$, $f_1=0.073(26)$, $f_2=0.0012(43)$, $f_3 = 0.0026(32)$ with a $\chi^2=1.25$. The large value of $\eta$ clearly establishes that this criticality is not captured by the chiral-Heisenberg theory, where $\eta$ computed numerically is approximately $0.45$. \cite{Otsuka:2015iba}

While the data so far has suggested that there is a second-order phase transition and that the $O(4)$ symmetry has been broken at couplings above $V_c/t$, and we also see more specifically that deep in the broken phase both the $\mathbb{Z}_2^{sc}$ and $SO(4)$ symmetries are broken because the correlations $C_U$ and $C_S$ saturate as seen in Figure \ref{fig:massive}, we also cannot completely rule out there being some intermediate phase where only the $SO(4)$ symmetry is broken, or only the $\mathbb{Z}_2^{sc}$ symmetry is broken. Using $V_c/t = 1.515$, which is the critical coupling we found from the scaling analysis for $C_S$, we partly address this question in Figure \ref{fig:O5} by plotting data at the critical coupling for $C_U$, which is odd under $\mathbb{Z}_2^{sc}$, but invariant under $SO(4)$. Here we see that the data fits well to a power law ($C_U \sim L^{-3.05(9}$) and does not appear to saturate. This is evidence against there being an intermediate phase that is $SO(4)$ symmetric but $\mathbb{Z}_2^{sc}$ broken.

\section{Conclusions}
We have introduced a family of interacting $N$-flavor Hamiltonians that is sign-problem-free, simulatable by the Hamiltonian fermion bag mathod, and possesses an $O(2N)$ flavor symmetry, just as free fermions do. We have focused specifically on the case where $N=2$ and discussed its $O(4)$ symmetry, which consists of spin and charge $SU(2)$ symmetries (forming an $SO(4)$ symmetry) as well as a spin charge flip symmetry $\mathbb{Z}^{sc}_2$. We have then shown numerical evidence for a second-order phase transition that breaks the model's $O(4)$ symmetry. Additionally we have presented numerical evidence that there is no intermediate phase that is $SO(4)$-symmetric but $\mathbb{Z}_2^{sc}$-symmetry broken. Using lattices up to $2304$ sites in size, we have computed critical exponents for the phase transition and have found them to be distinct to those in the chiral Heisenburg universality class, making this a distinct spin-charge flip symmetric fixed point in $2+1d$.

\section*{Acknowledgements}
This work was done in collaboration with Shailesh Chandrasekharan, Ribhu Kaul, and Hanqing Liu. It used computational resources provided by the Extreme Science and Engineering Discovery Environment (XSEDE) \cite{xsede}. Research at the Perimeter Institute is supported
in part by the Government of Canada through the Department of Innovation, Science and Economic Development and by the Province of Ontario through the Ministry of Colleges and Universities.

\end{document}